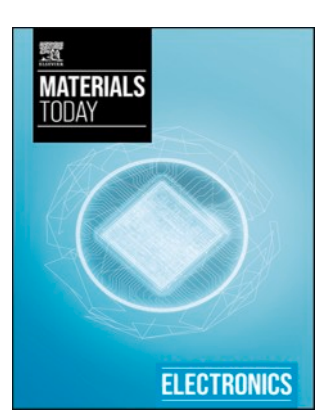

Full Length Article

# Stabilized biskyrmion states in annealed CoFeB bilayer with different interfaces

Warda Al Saidi [a,1], Selma Amara [b], Myo T. Zar Myint [a], Salim Al Harthi [a], Gianluca Setti [b], Rachid Sbiaa [a,*]

[a] Department of Physics, College of Science, Sultan Qaboos University, P.O. Box 36, PC 123, Muscat, Oman
[b] Division of Computer, Electrical and Mathematical Sciences and Engineering, King Abdullah University of Science and Technology (KAUST), Thuwal, Saudi Arabia

ARTICLE INFO

ABSTRACT

This study investigates the stability of skyrmions and biskyrmions in perpendicular magnetic tunneling junctions with a thick CoFeB/Ta/CoFeB free layer. The samples showed a magnetoresistance of ~ 41 % when annealed at 230 °C. Magnetic force microscopy revealed the existence of skyrmions and biskyrmions at room temperature in the as-deposited state and under an external magnetic field. Annealing at 330 °C enhanced interfacial Dzyaloshinskii-Moriya interaction (DMI) and crystallinity, enabling the spontaneous coexistence of these topological structures. Micromagnetic simulations explored the interplay between DMI strength, sign, and skyrmion chirality. Skyrmions exhibited repulsive interactions while biskyrmions displayed attractive interactions due to the difference in helicities. The study highlights the influence of multilayer structure and varying Ta layer thicknesses on the DMI chirality, which modulates the formation of complex spin textures. These results provide an understanding of skyrmion and biskyrmion dynamics and their potential for spintronic applications, including racetrack memory and data storage technologies.

## 1. Introduction

Since the experimental discovery of magnetic skyrmions in 2009, the field of magnetic topological structures has experienced a surge in interest and rapid development [1–10]. Skyrmions, as topologically nontrivial spin textures, exhibit nanoscale dimensions, remarkable stability, and the ability to be manipulated with low driving current densities [11]. These properties have positioned skyrmions at the forefront of spintronics research. Over the past decade, a large number of magnetic spin configurations have been identified, expanding beyond the archetypal skyrmion to include antiskyrmions, merons, antimerons, biskyrmions, and even triskyrmions [12–14]. Each of these spin structures is characterized by distinct topological charges, ranging from $Q = 0$ for skyrmionium to $Q = \pm 1$ for Néel and Bloch skyrmions to $Q = \pm 2$ for biskyrmions and fractional values for merons ($Q = \pm\frac{1}{2}$) [15]. The topological charge, determined by spin chirality, encapsulates key details about their structure and distinguishes, for example, skyrmions from antiskyrmions or merons from antimerons. In addition to these spin configurations, more complex entities such as skyrmion tubes [16,17], magnetic flows, and hopfions [18,19] further enrich the skyrmion family. The parameters of vorticity $\eta$ and helicity $\gamma$ provide additional means to classify these textures [20].

Among these fascinating spin textures, biskyrmions hold particular significance. These are composed of two partially overlapping skyrmions with opposite helicities, resulting in a total topological charge of $Q = \pm 2$ [21]. Despite their merged nature, each skyrmion within a biskyrmion retains its individual identity. This partial overlap characterizes the biskyrmion as a stable bound state, with each skyrmion contributing a charge of $Q = \pm 1$. Magnetic dipolar interactions often stabilize a biskyrmion, unlike Dzyaloshinskii-Moriya interactions (DMIs) that stabilize a skyrmion. Dipolar interactions induce inter-skyrmion forces, like Lennard-Jones potentials, which favor transitioning from a skyrmion state to a biskyrmion state. Furthermore, the centrosymmetric nature of host materials plays a crucial role in their stabilization.

The first experimental observation of biskyrmions was reported in the centrosymmetric layered manganite material known for its colossal magnetoresistance [22]. Subsequent investigations have demonstrated the formation of high-density biskyrmion lattices in various materials, including $Ho(Co,Fe)_3$ [2], MnNiGa [23–26], rare-earth substituted alloys [27] and Fe-Gd thin films [28]. These studies revealed that skyrmions could persist across a wide temperature range, sometimes even at room temperature and under zero external magnetic field. In this study, we report on forming biskyrmions in a thick CoFeB layer. Our results reveal the formation of both skyrmions and biskyrmions in these thin films.

* Corresponding author.
*E-mail address:* rachid@squ.edu.om (R. Sbiaa).

[1] Permanent address: Physics Unit, Department of Sciences, College of Applied Sciences and Pharmacy, University of Technology and Applied Sciences, Muscat, Oman.

## 2. Results and discussion

The investigated structure is a CoFeB(1.2)/Ta(0.3)/CoFeB(1.2) deposited on MgO(1.5) itself on an antiferromagnetically coupled structure. This later consists of a $[Co(0.5)/Pt(0.6)]_{\times 6}$/Co(0.3)/Ru(0.4)/$[Co(0.5)/Pt(0.6)]_{\times 2}$/Co(0.3)/Ta(0.25)/CoFeB(0.85), where the individual layer thicknesses are denoted in nanometers as shown schematically in Fig. 1(a). Without 0.3 nm-thick Ta, the CoFeB free layer with 2.4 nm will have an effective in-plane magnetic anisotropy due to the reduced interface anisotropy [29,30]. The insertion of the Ta spacer layer in magnetic tunnel junctions (MTJs) enhances the thermal stability and efficiency of current switching in perpendicular magnetic easy axis (p-MTJs). Prior to the deposition of the whole MTJ structure, the thickness of Ta separation layer was investigated. Specifically, a 1.0 nm-thick Ta spacer results in independent switching of CoFeB layers, whereas a thinner Ta spacer leads to a strong ferromagnetic coupling and the out-of-plane magnetization is prevented. Further studies suggest that the CoFeB layer thickness can be increased by increasing the Ta thickness, with a 0.5 nm Ta insertion resulting in a doubling of the effective thickness of the CoFeB layer, significantly enhancing the perpendicular anisotropy [30]. The structure also features a 1.5 nm MgO tunnel barrier, positioned on the top of a CoFeB (0.85 nm) layer, contributing to the overall performance of the MTJ. The whole stack was deposited at room temperature on oxidized silicon wafers under a base pressure of $<8 \times 10^{-9}$ mbar. Notably, using a Ta capping layer in the top CoFeB layer plays a critical role in enhancing the tunnel magnetoresistance (TMR) ratio and improving the temperature stability by minimizing atomic diffusion. The Ta capping layer also facilitates the absorption of boron atoms from the CoFeB layer, crucial for establishing strong interfacial anisotropy at the CoFeB/MgO interface. Additionally, studies have shown that a Ru capping layer can improve interfacial anisotropy and reduce the damping constant, thereby enhancing the thermal stability of MTJ free layers. While both Ru and Ta capping layers yield similar TMR ratios, MTJs with Ru exhibit higher coercivity and better thermal stability [31]. The samples underwent a post-annealing at temperatures of 230 °C and 330 °C for 30 min in a high-vacuum environment with a heating rate of 10 °C/min. The MTJ nanopillars with diameters ranging from 1 μm to 2 μm were fabricated by photolithography and ion milling as shown in Fig. 1(b). The magnified view in Fig. 1(c) highlights the structure of the top and bottom electrodes.

Magneto-optical Kerr hysteresis loops in polar geometry provide a clear signature of perpendicular magnetic anisotropy (PMA) in CoFeB-based multilayers as shown in Fig. 1(d). A nearly square out-of-plane loop with steep sides and high remanent magnetization denotes a robust perpendicular easy axis, as the magnetization remains largely up or down at remanence state (no applied field). In contrast, a tilted or sheared loop indicates that the magnetization is partially canted or forming multi-domain states at remanence, reflecting weaker effective PMA [32]. Thus, loop squareness serves as sensitive proxies for interfacial anisotropy strength in these ultrathin CoFeB/MgO/Ta films. Annealing at 230 °C yields a loop with relatively high squareness and minimal tilt, consistent with a strong interfacial PMA developed via CoFeB crystallization at the MgO interface. In contrast, the sample annealed at 330 °C exhibits a more tilted hysteresis loop with lower remanence, indicating a reduction in perpendicular anisotropy compared to the 230 °C case. Notably, the coercive field remains of the same order for both annealing conditions investigated. It has been reported that excessive thermal treatment can induce diffusion of Ta into the CoFeB layer, which degrades the sharp interfaces responsible for PMA [33]. The intermixing is also likely to weaken the interfacial DMI, since it originates from the heavy-metal/ferromagnet interface's broken symmetry and strong spin–orbit coupling. Both PMA and DMI are key parameters for stabilizing chiral nanoscale spin textures where a balanced combination of high PMA and sufficient DMI is required to favour stable skyrmions or biskyrmions [34]. Strong PMA helps confine the skyrmion core while DMI endows domain walls with a fixed chirality and lowers their energy [35].

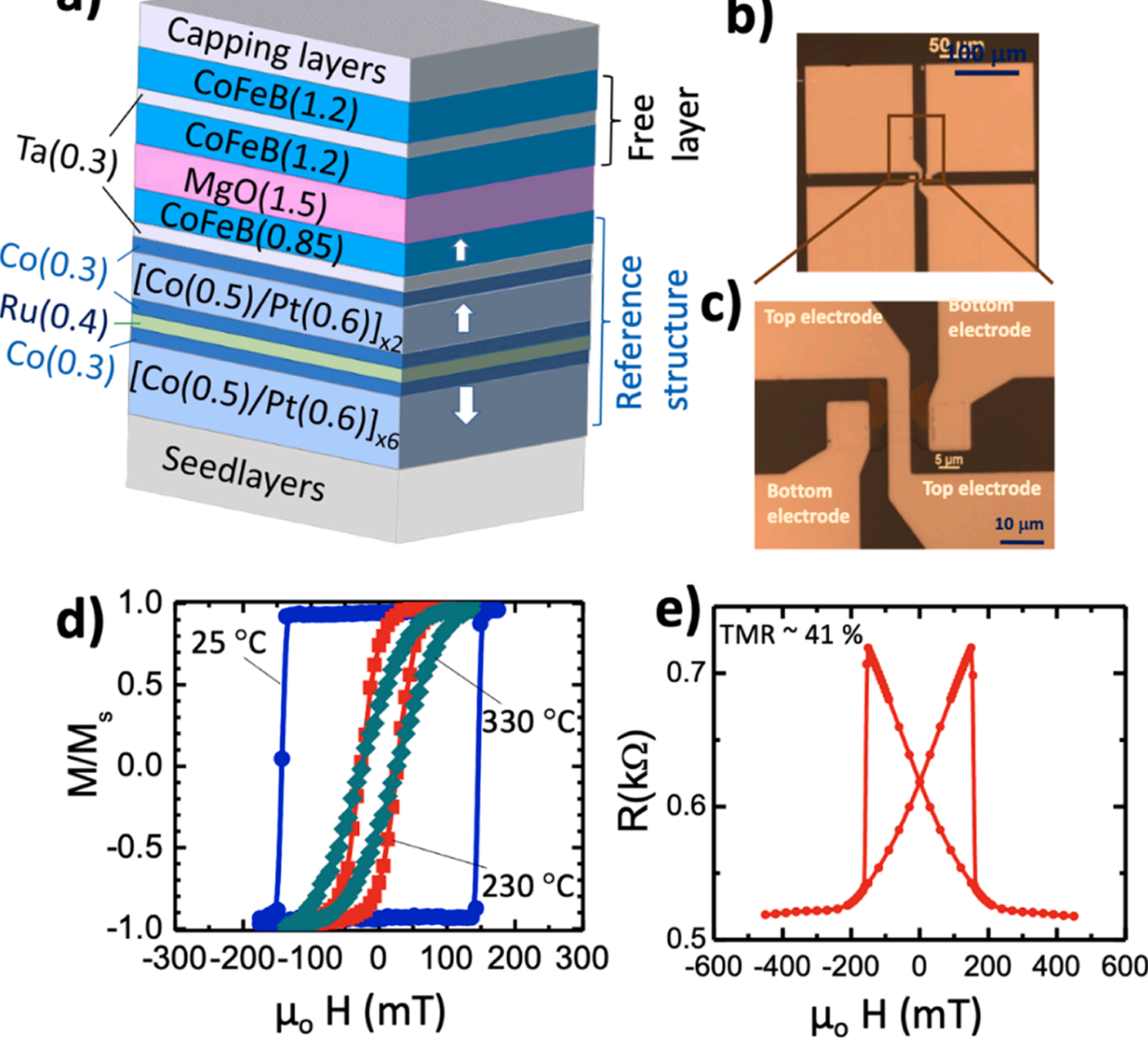

**Fig. 1.** Schematic representation of the MTJ where the free layer is made of CoFeB/Ta/CoFeB stack. Panel (b) depicts the MTJ nanopillars with diameters ranging from 1 to 2 μm, and panel (c) provides a magnified view of the top and bottom electrode structure. (d) Hysteresis loop of the out-of-plane magnetization component of the free layer at as-deposited state and after annealed at elevated and (e) Tunneling magnetoresistance properties of the MTJ annealed at 230 °C.

To confirm the electrical functionality of the MTJ stack, TMR was measured at room temperature on a single nanopillar device. Fig. 1(e) presents the junction resistance $R$ as a function of the magnetic field $H$ for a 1.3 μm-diameter MTJ annealed at 230 °C. The devices were characterized using a four-point contact geometry showing 41 % TMR. Although the primary focus of this work lies in imaging and understanding biskyrmion formation, the TMR measurement indicates that the magnetic configuration in the device is well-defined and compatible with practical spintronic applications.

Magnetic force microscopy (MFM) was utilized to examine the formation and stability of skyrmions in the as-deposited state and following annealing. The Co/Pt multilayers and CoFeB/MgO interfaces exhibit pronounced PMA, which is essential for stabilizing the out-of-plane magnetization; a key condition for skyrmion formation. Although annealing can further enhance PMA, the as-deposited state already demonstrates sufficient anisotropy to support skyrmion nucleation. Fig. 2 highlights the presence of both magnetic domains and skyrmions in the as-deposited state at room temperature. The $z$-mode MFM images in Fig. 2(a) clearly identify skyrmions coexisting with magnetic domains, while the corresponding $\Delta f$-mode image in Fig. 2(b) provides complementary confirmation of their topological characteristics. Similar observations were made in other regions of the sample, where skyrmions and domain walls were identified in the $z$-mode (Fig. 2(c)) and $\Delta f$-mode (Fig. 2(d)) images. These findings emphasize the robustness of skyrmion formation in the as-deposited state, supported by the inherent properties of the multilayer structure. The coexistence of magnetic domains in the as-grown state of the sample at an annealing temperature of 230 °C was observed; no skyrmions were detected. Annealing at 230 °C enhances the crystallinity in the material, but it appears insufficient to induce the necessary conditions for skyrmion formation. The coexistence of magnetic domains across the thin film suggests that PMA in the system is strong enough to confine the magnetization out-of-plane, yet not optimized for skyrmion stabilization without applying an external perturbation. Upon applying a magnetic field of $H$ = 15 mT, however, skyrmions and biskyrmions emerged across the thin film. These topological structures were observed in both $z$-mode (Fig. 3(a)) and $\Delta f$-mode (Fig. 3(b)) of MFM, with magnified regions showing more details in Fig. 3(c) for $z$-mode and Fig. 3(d) for $\Delta f$-mode. Applying an external field effectively reduces the energy barrier for skyrmion formation by aligning the magnetization and promoting a localized energy minimum that supports the skyrmion structure. The combination of strong PMA and DMI originating from the multilayer stack, facilitates the stabilization of these topological states under the applied field.

The frequency shift mode $\Delta f$ is highly sensitive to weak magnetic fields and is directly related to the force gradient between the magnetic tip and the magnetic textures. In contrast, in the height shift mode $\Delta z$, the magnetic force is recorded at a constant height. The imaging scan is performed line by line after two passes. In the first one, surface topography (physical textures) is recorded, then in the second pass, the height is fixed and the magnetic response is captured. There is a high possibility of affecting the magnetic textures during the first scan. The image taken by $\Delta f$-mode represents the derivative of the one taken by $\Delta z$-mode but with much high sensitivity. The $\Delta f$-mode is preferable for very small magnetic textures that could be affected by an external magnetic field, as the scan can be carried out at a large distance from the sample surface.

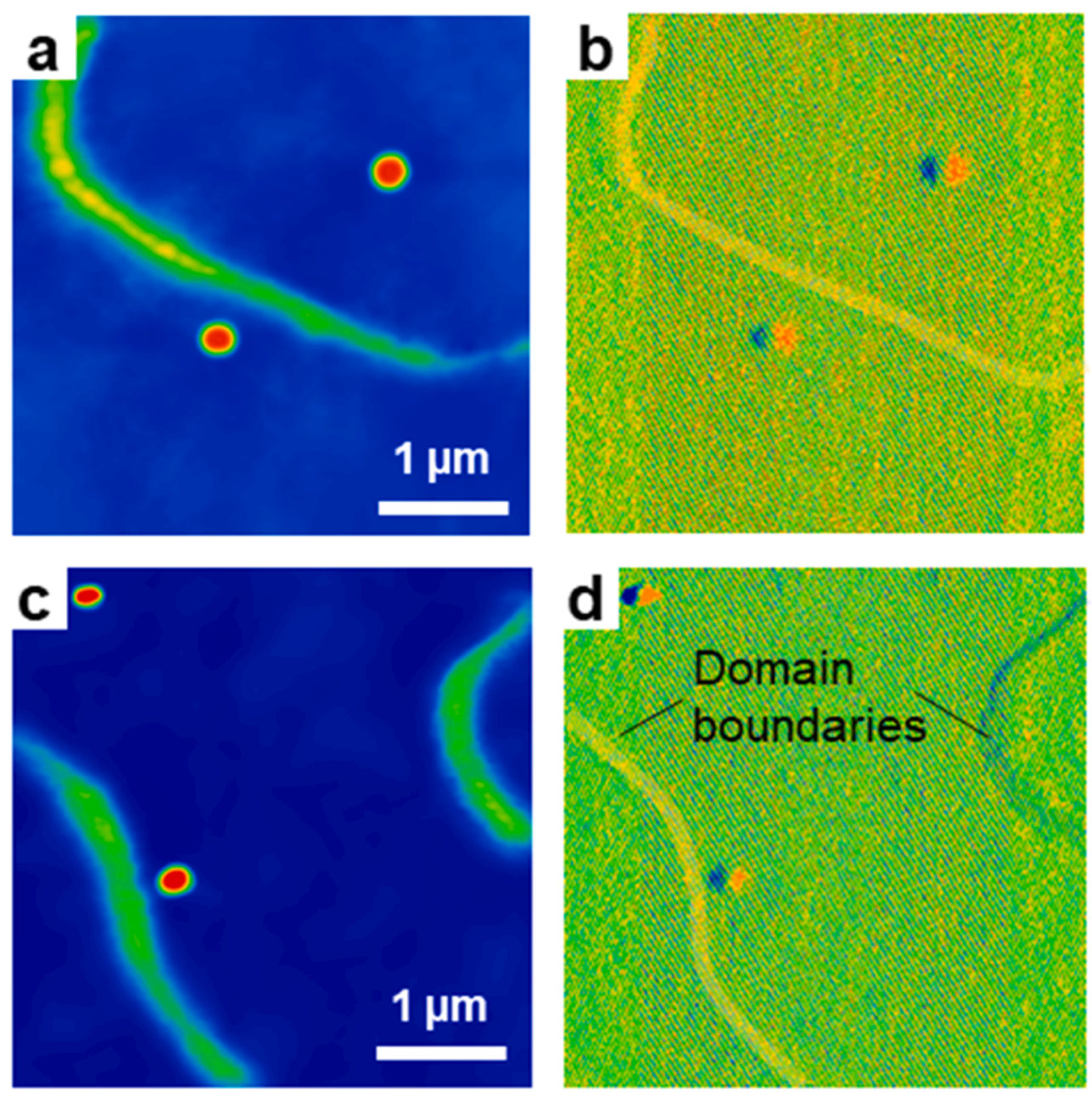

**Fig. 2.** Skyrmion Formation and Stability in the As-Deposited State. (a) $z$-mode Magnetic force microscopy (MFM) image showing skyrmions coexisting with magnetic domains at room temperature. (b) Corresponding $\Delta f$-mode MFM image providing complementary confirmation of the skyrmions' topological characteristics. (c) $z$-mode MFM image from another region of the sample showing skyrmions and domain walls. (d) Corresponding $\Delta f$-mode MFM image of the same region, confirming the presence of skyrmions and domain walls.

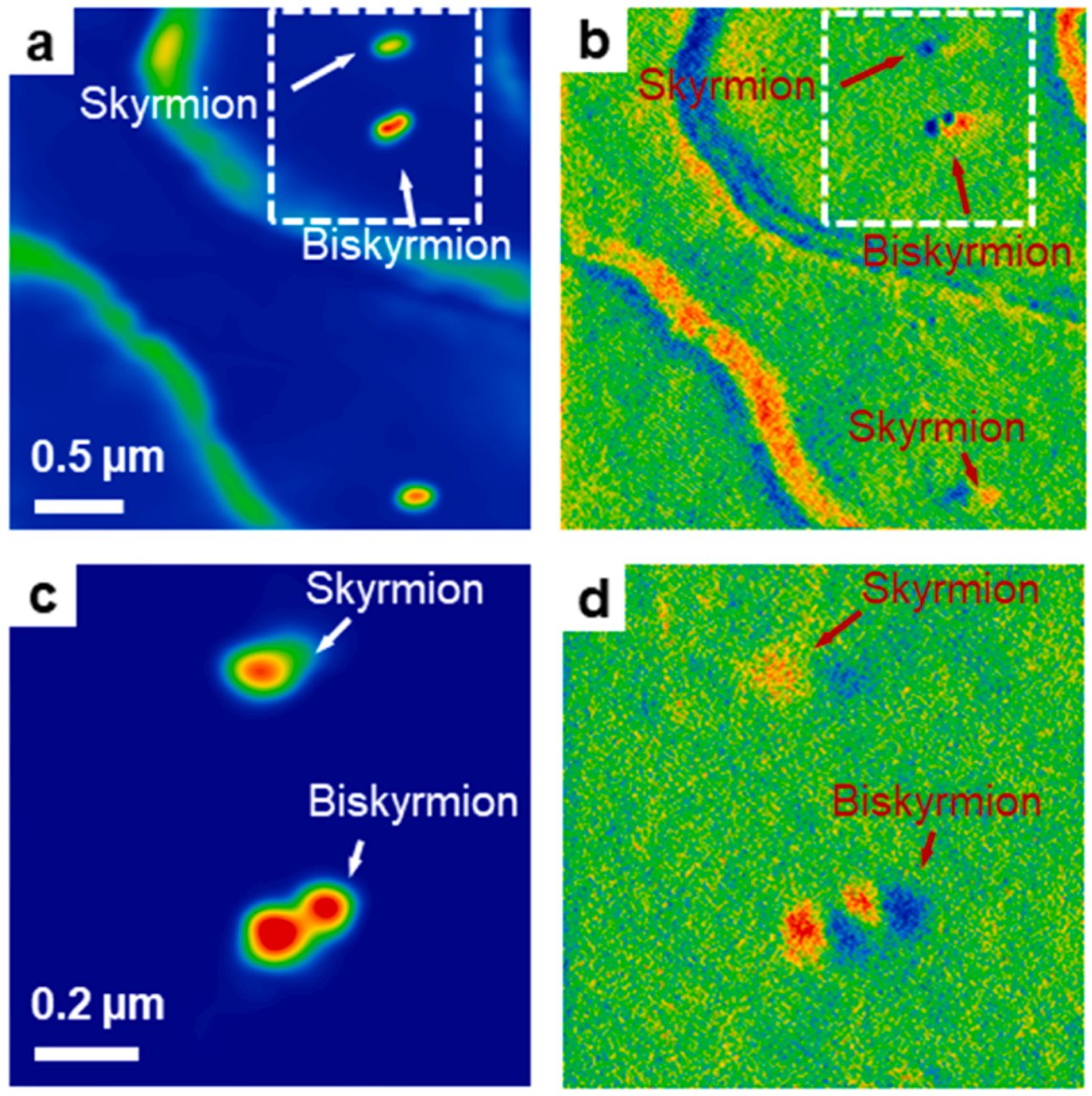

**Fig. 3.** Skyrmion and biskyrmion formation at 230 °C annealing temperature. (a) Skyrmions and biskyrmions observed in the $z$-mode of MFM under a magnetic field of $H$ = 15mT. (b) Corresponding image of skyrmions and biskyrmions in $\Delta f$-mode. (c) Magnified view of the skyrmion and biskyrmion structures in the $z$-mode. (d) Magnified view of the same textures in $\Delta f$-mode.

The coexistence of skyrmions and biskyrmions reflects the interplay between the intrinsic properties of the material and the external field, enabling diverse magnetic textures, as seen in similar studies of thin films with tailored DMI interactions [36]. The structural design of the multilayer thin film, comprising Co/Pt, CoFeB, Ta, Ru, and MgO layers, introduces a complex DMI landscape due to the varying thicknesses of Ta. It is possible for the structure to exhibit both positive and negative DMI contributions due to these varying thicknesses [37,38]. It is important to note that DMI is not necessarily spatially uniform across a real magnetic sample. The magnitude and even sign reversals of DMI can be spatially influenced by variations in local strain, interface roughness, or compositional gradients at asymmetric heavy metal/ferromagnet interfaces, which can affect the spin texture stability and chirality [39]. In thinner layers, the DMI at the interface tends to dominate and may favor a specific chirality, whereas in thicker layers, contributions from multiple interfaces might alter the net DMI and potentially reverse its sign. The interfaces are known to generate significant DMI, but the contributions of these interfaces can differ in magnitude and sign depending on layer thickness and interfacial properties [40,41]. Different Ta thicknesses result in varying degrees of spin-orbit coupling and interfacial symmetry breaking. The thinner Ta layer (0.25 nm) produces a weaker DMI due to reduced spin-orbit coupling, while the thicker Ta layer (1 nm) may generate a stronger or differently oriented DMI due to enhanced spin-orbit interaction and changes in the interfacial spin configuration [40]. As a result, the asymmetry in the multilayer structure, caused by the combination of different Ta thicknesses, can lead to competing interfacial DMI contributions. This competition may stabilize complex spin textures, including the coexistence of skyrmions with opposite chiralities, stripe domains, or even biskyrmions, as observed in engineered multilayer systems with tailored DMI gradients [42]. The dependence of DMI on Ta layer thickness has been well-documented, with studies demonstrating that the DMI strength and sign are highly sensitive to the thickness of the Ta layer. For example, thinner Ta layers (e.g., Ta 0.25 nm) tend to favor a particular chirality, while thicker Ta layers (e.g., 1 nm thick-Ta) can change the DMI chirality due to altered electronic properties and spin-orbit effects [36]. Furthermore, the sign and strength of the interfacial DMI can be influenced by material composition, stack order, and interface quality, as demonstrated by the observed dependence of DMI strength with ferromagnetic layer thickness [40]. At an annealing temperature of 330 °C, skyrmions and biskyrmions coexist across the thin film in its as grown state; i.e. without applying a magnetic field. This elevated annealing temperature significantly improves film crystallinity, ordering, and overall material properties. These changes likely enhance the interfacial DMI, optimizing the conditions necessary for skyrmion stabilization without the need of an external magnetic field. The higher temperature facilitates the alignment of magnetic moments at the interfaces, which enhances the intrinsic properties of the film, making it more favorable for the spontaneous formation of skyrmions and biskyrmions. The improved film quality at 330 °C enables the coexistence of these topological textures in the absence of external perturbations. As shown in Fig. 4, panel (a) presents the existence of biskyrmions in the $z$-mode, while panel (4b) displays the biskyrmion structure in the $\Delta f$-mode. Additionally, panel (4c) reveals both skyrmions and biskyrmions in the height-mode, with panel (4d) showing the same textures in the $\Delta f$-mode. These observations demonstrate the spontaneous emergence of skyrmions and biskyrmions under the optimized conditions created by the higher annealing temperature.

To investigate the effect of skyrmion chirality on the configuration of repulsion and biskyrmion states, micromagnetic simulations were performed using the MuMax3 software package [43]. The simulations are based on solving the Landau–Lilshitz–Gilbert (LLG) equation for the time evolution of the magnetization. In its continuous form, the LLG equation can be written as

$$\frac{d\boldsymbol{m}}{dt} = \gamma_o \boldsymbol{m} \times \boldsymbol{H}_{\text{eff}} \; + \; \alpha \, \boldsymbol{m} \times \frac{d\boldsymbol{m}}{dt} \tag{1}$$

where $\boldsymbol{m}(\boldsymbol{r}, t) = \boldsymbol{M}/M_s$ is the magnetization vector unit which is dependent on the position and time. Here $\gamma_o \; = |\gamma|\mu_o$is the gyromagnetic ratio (with $\gamma$ the electron gyromagnetic factor and $\mu_o$ the vacuum permeability) and $\alpha$ is the Gilbert damping constant. The effective field $\boldsymbol{H}_{\text{eff}} = -\frac{1}{\mu_o M_s} \; \frac{\delta E}{\delta \boldsymbol{M}}$ is derived from the total magnetic energy $E$ of the system, and includes contributions from exchange interaction, magnetic anisotropy, magnetostatic fields, the interfacial DMI, and any applied external field. On the right side of Eq. (1), the first term represents the precession of $\boldsymbol{m}$ around $\boldsymbol{H}_{\text{eff}}$, while the second term is the damping that drives the system towards local energy minima (dissipative dynamics).

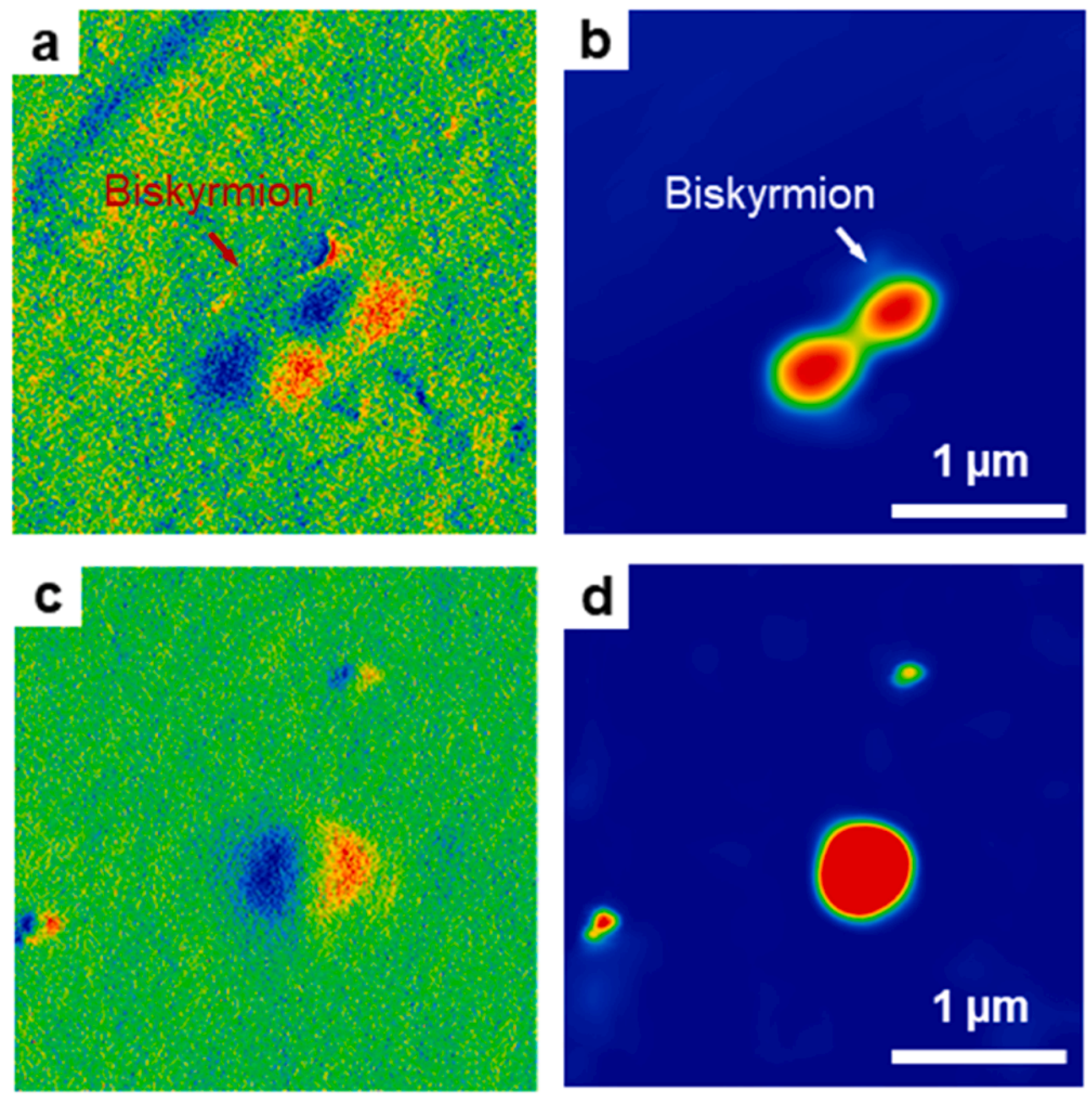

**Fig. 4.** Coexistence of skyrmions and biskyrmions in the thin film annealed at 330 °C. Panel (a) illustrates the presence of biskyrmions observed in $\Delta f$-mode, (b) the biskyrmion structure captured in the $z$-mode. (c) skyrmions and biskyrmions in $\Delta f$-mode, with panel (d) presenting the corresponding textures in the $z$-mode.

The material parameters for the simulations were set as follows: a saturation magnetization $M_s = 900$ kA/m, and exchange stiffness $A_{\text{ex}}$ $=1.3 \times 10^{-11}$ J/m. The Gilbert damping parameter was set to $\alpha = 0.1$ to facilitate numerical convergence; this relatively high damping accelerates the relaxation of the magnetization configuration into equilibrium but does not appreciably alter the static domain structure solutions. The system magnetic anisotropy was modeled using a uniaxial perpendicular anisotropy constant of $0.6 \times 10^6$ J/m $^3$.

The interfacial DMI was included in the model, as our system breaks inversion symmetry at the interfaces and exhibits chiral spin textures. The interfacial DMI is implemented in the simulation as an effective field term derived from the antisymmetric exchange interaction, specifically adapted for thin-film geometries where broken inversion symmetry at interfaces dominates. The effective DMI field is expressed as [44,45]

$$\boldsymbol{H}_{\text{DMI}} = \frac{2D}{\mu_o M_s} \left( \frac{\partial m_z}{\partial x}, \frac{\partial m_z}{\partial y}, -\frac{\partial m_x}{\partial x} - \frac{\partial m_y}{\partial y} \right) \tag{2}$$

where $D$ is the interfacial DMI constant (in units of J/m$^2$). This form of DMI favors Néel-type domain walls and skyrmions, imparting a fixed chirality (handedness) to the rotation of spins across the wall. The sign of $D$ dictates the chirality. We examined values of $|D|$ on the order of 1 mJ/m$^2$ in our simulations, which is a realistic magnitude for CoFeB-based interfacial systems [46]. In particular, simulations were carried out with $D = +1.0$ mJ/m$^2$ and $D = -1.0$ mJ/m$^2$ to represent the two opposite chiral orientations. If $D$ is set to 0, no stable skyrmion state is found, skyrmion stability relies on the presence of DMI or similar chiral interactions [47,48].

All simulations were performed on a discretized mesh of 256 × 256 × 1, with a cell size 1 nm$^3$. This high in-plane resolution (1 nm) is necessary to adequately resolve the domain wall width and skyrmion core structure, which for our material parameters is on the order of a few nanometers. The simulated sample was chosen to be large enough to accommodate multiple skyrmions and to minimize finite-size effects (open boundary conditions were applied in-plane, allowing spins at the boundary to align with the nearest neighbors with no artificial pinning). To specifically explore the interaction of skyrmions with opposite chirality (the scenario relevant to biskyrmion formation in our bilayer), we took the following approach: We first simulated two skyrmions in a single layer with a uniform DMI (D constant) to see how they interact. A negative DMI of $-1$ mJ/m$^2$ induced one chirality, while a positive DMI value would promote the opposite chirality.

Two proximate Néel skyrmions exhibit a repulsive interaction and thus tend to repeal each other to maximize their separation, as they have like topological charge and identical wall rotation. This approach allows for studying the transition from skyrmion repulsion to biskyrmion formation as a function of DMI sign and strength. Skyrmions, characterized by a topological charge of $Q = 1$, exhibit repulsive interactions due to their similar charge and chirality. This effective skyrmion-skyrmion repulsion decays exponentially with increasing separation distance [49]. The repulsion ensures that skyrmions prefer to maintain a distance from each other, which has significant implications for skyrmion-based racetrack memory devices. Reliable and practicable initial spacing, such as $d_i > 62$ nm, is necessary for properly writing and reading consecutive skyrmionic bits, as validated by Sampaio et al. [50]. These spacings mitigate the influence of mutual repulsion on skyrmionic bits, enabling stable operation of RM systems [49,51]. A critical separation $d$ has been observed, below which skyrmion pairs collapse, leaving a single skyrmion with $Q = 1$. As $d$ decreases, the system total energy increases, accompanied by a reduction in skyrmion size λ. This phenomenon highlights the dynamic nature of skyrmion interactions, as reported by Capic et al. [52]. Finally, to mimic a bilayer where each layer favors an opposite chirality, we performed a thought-experiment by considering one skyrmion with positive and a second skyrmion with negative in proximity. In this mixed-chirality scenario, we observed an attractive interaction between the two skyrmions moving towards each other and eventually merged into a bound pair, i.e., a biskyrmion. The attraction arises because in each skyrmion chiral domain wall spins are complementary to the other, allowing them to share a partial domain wall and lower the overall energy. This result from the simulation directly supports our experimental claim that the two CoFeB sub-layers host skyrmions of opposite chirality which couple into a biskyrmion topological texture.

Fig. 5 has been annotated to indicate the chirality in each panel. The left column shows snapshots for $D > 0$ where both skyrmions are left-handed Néel type and repel each other). The right column shows the formation of a biskyrmion in the mixed-chirality case where one left-handed Néel skyrmion is attracting the other right-handed type. For visualization, spins are color-coded by their out-of-plane component $m_z$, with blue indicating magnetization in the up direction (positive $m_z$) and red indicating magnetization down (negative $m_z$). Arrows on the spins depict the orientation, highlighting the rotation. The micromagnetic modeling strongly supports the interpretation that opposite DMI signs (originating from the distinct CoFeB interfaces in the experiment) lead to opposite skyrmion chiralities, which in turn enable the formation of biskyrmions through an attractive interaction.

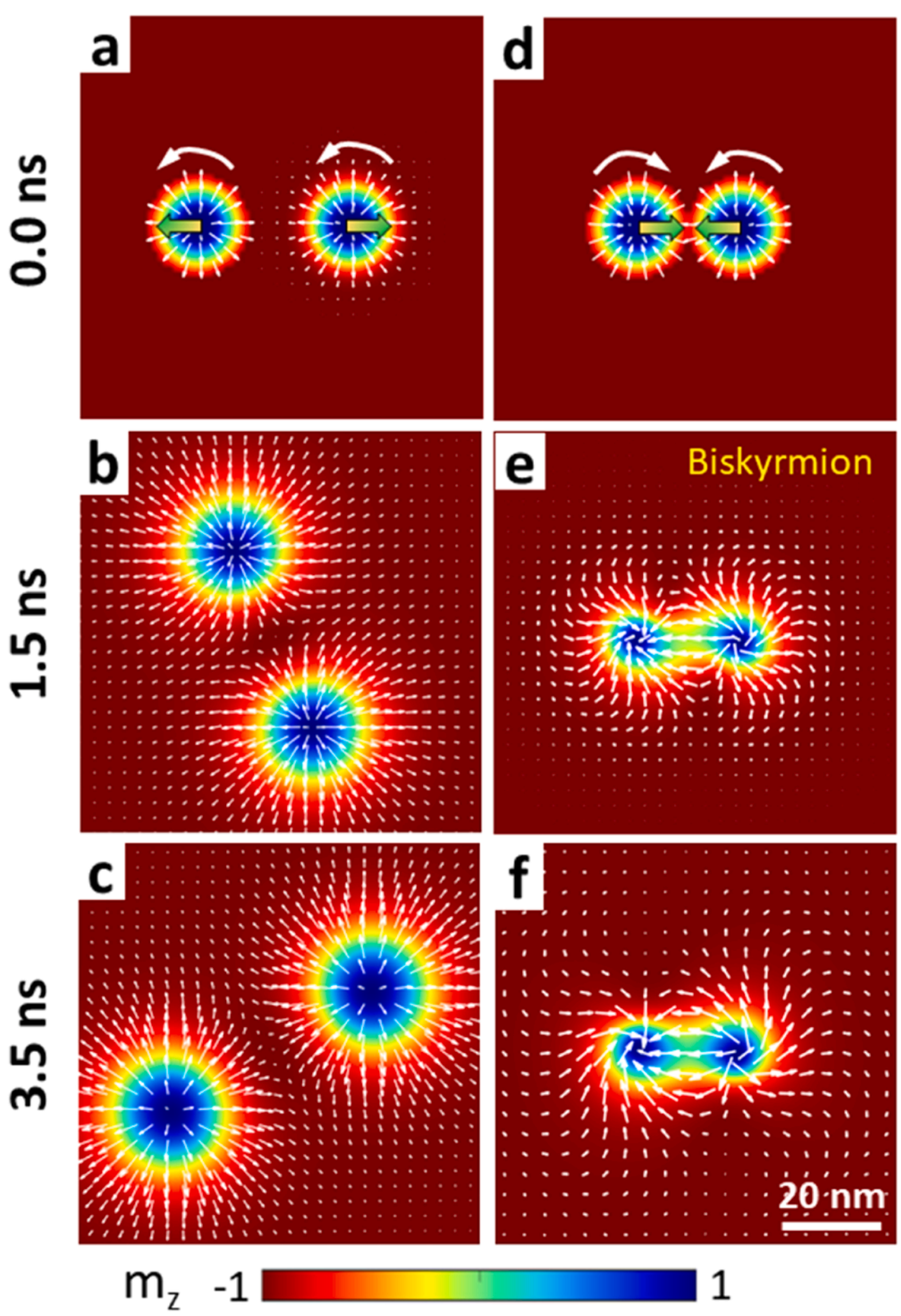

**Fig. 5.** Spins Configurations of two Skyrmions $m_z$ Left Column: Néel skyrmions with the same chirality (γ = 0), exhibit repulsive interactions. The spins configurations show outward rotation, resulting in skyrmions moving away from each other. Right Column: Formation of a biskyrmion, characterized by opposite chirality of the constituent skyrmions. The spins in this configuration are complementary, resulting in an attractive interaction stabilizing the biskyrmion structure. The color map represents the out-of-plane magnetization component ($m_z$), with red and blue indicating positive and negative values, respectively.

## 3. Conclusion and outlook

This study explores the formation of skyrmions and biskyrmions in a double CoFeB free layer with perpendicular anisotropy. The electrical characterization of the junction showed a magnetoresistance of about 41 % after annealing at 230 °C. The magnetic force microscopy revealed the presence of skyrmions and biskyrmions. For 330 °C annealing temperature, skyrmions and biskyrmions spontaneously coexisted without applying a magnetic field. The study shows that the two CoFeB layers composing the free layer generate skyrmions with different chiralities due to different interfaces leading to biskyrmions. The results were supported by micromagnetic simulations, which confirm that biskyrmions exhibit attractive interactions due to their opposite helicities. The findings provide insight into skyrmion dynamics in magnetic tunnelling junctions and could be applied to functional spintronic devices.

## CRediT authorship contribution statement

**Warda Al Saidi:** Writing – review & editing, Investigation. **Selma**

**Amara:** Writing – review & editing, Investigation. **Myo T. Zar Myint:** Formal analysis. **Salim Al Harthi:** Resources. **Gianluca Setti:** Investigation. **Rachid Sbiaa:** Writing – review & editing, Supervision, Conceptualization.

## Declaration of competing interest

The authors declare that this work has not been published previously, that it is not under consideration for publication elsewhere, that its publication is approved by all authors and tacitly or explicitly by the responsible authorities where the work was carried out, and that, if accepted, it will not be published elsewhere in the same form, in English or in any other language, including electronically without the written consent of the copyright-holder.

## Data availability

Data will be made available on request.